# Design of Planar and Conformal, Passive, Lossless Metasurfaces that Beamform

Jordan Budhu, *Member, IEEE*, Luke Szymanski, *Member, IEEE*, and Anthony Grbic, *Fellow, IEEE*

*Abstract*—A general technique for synthesizing both planar and conformal beamforming metasurfaces is presented that utilizes full-wave modeling techniques and rapid optimization methods. The synthesized metasurfaces consist of a patterned metallic cladding supported by a finite-size grounded dielectric substrate. The metasurfaces are modeled using integral equations which accurately account for mutual coupling and the metasurface's finite dimensions. The synthesis technique consists of three-phases: a direct solve phase to obtain an initial metasurface design with complex-valued impedances satisfying the desired far-field beam specifications, a subsequent optimization phase that converts the complex-valued impedances to purely reactive ones, and a final patterning phase to realize the purely reactive impedances as a patterned metallic cladding. The optimization phase introduces surface waves which facilitate passivity. The metasurface is optimized using gradient descent with a semi-analytic gradient obtained using the adjoint variable method. Three examples are presented: a low-profile directly-fed metasurface antenna with near perfect aperture efficiency, a scanned-beam reflectarray design with controlled sidelobes, and a conformal metasurface reflectarray. The far-field and near-field performance of the metasurfaces are verified and the bandwidth and loss tolerance of the metasurfaces are investigated.

## I. INTRODUCTION

METASURFACES are thin subwavelength textured surfaces (see Fig. 1). They are often modeled using idealized bianisotropic interfaces known as generalized sheet transition conditions or GSTC's [1]. The texturing of the surface can be modeled using electric and magnetic admittances/impedances and magnetoelectric coupling parameters, which form proportionality constants between induced surface currents and electric and magnetic fields tangential to the surface. By introducing surface waves onto these metasurfaces, they can be used for far-field beamforming in a completely passive and lossless manner requiring no local loss or gain, only simple patterned metallic claddings on RF substrates [2]–[23]. The surface waves supported by beamforming metasurfaces reshape the local power density by redistributing power transversally along the metasurface from places where local loss is needed to places where local gain is needed, a concept first introduced in [9]. For example, the metasurfaces in [3]–[5], [8]–[23] use surface waves to control either the near or far-fields in a passive and lossless manner.

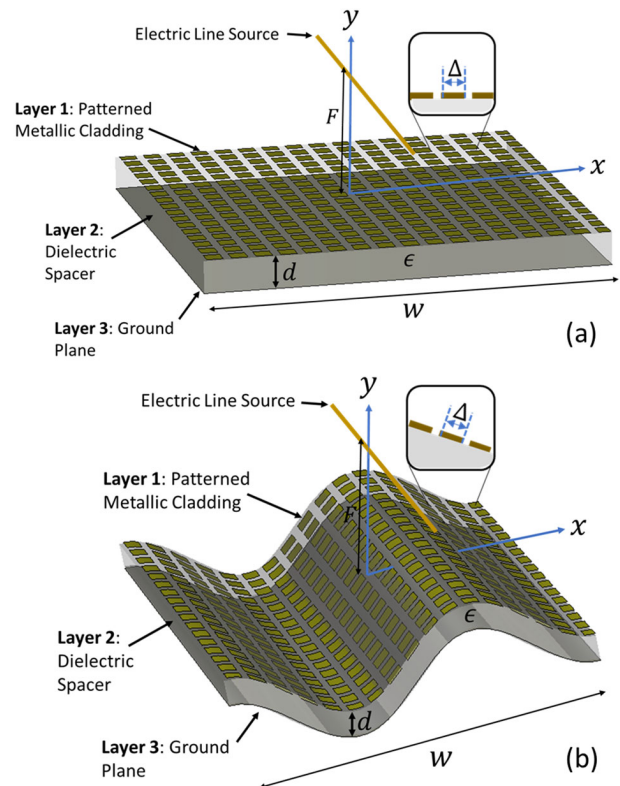

Fig. 1. Metasurface geometry. The metasurface is infinite and invariant in the $z$-direction and finite and spatially variant in the $x$-direction. The metasurface consists of three layers: a periodic or aperiodic patterned metallic cladding layer (layer 1) separated from a ground plane (layer 3) by a dielectric spacer (layer 2). (a) Planar metasurface beamformer. (b) Conformal metasurface beamformer.

When employing surface waves in metasurface beamformer design, however, care must be taken to ensure that the realized prototype can support the same surface waves as the homogenized sheets used in design. For example, a homogenized sheet may theoretically be able to support a surface wave with a large

Jordan Budhu, Luke Szymanski, and Anthony Grbic are with the Electrical and Computer Engineering Department, University of Michigan, Ann Arbor, MI 48109 USA (e-mail: jbudhu@umich.edu, ljszym@umich.edu, agrbic@umich.edu).

TABLE I
CATEGORIZATION OF RECENT SCIENTIFIC WORKS ON METASURFACE
BEAMFORMER DESIGN INVOLVING OPTIMIZATION OF BOUNDARIES

| | Opt. over $E$ | Opt. over $\eta$ | Opt. over $J$ |
|---|---|---|---|
| References | [3], [4], [10], [11], [17], [21] | [5], [7], [12]–[16], [23] | [2], [14], [19], [25] |

Note: Opt. stands for Optimization.

tangential wavenumber, while a discretized, patterned metallic cladding has a tangential wavenumber cutoff. Further, a realizable metasurface consisting of a patterned metallic cladding (electric impedance sheet) over a conductor-backed RF substrate has a finite thickness. The finite thickness introduces spatial dispersion which gives rise to a set of surface wave modes different from those supported by an idealized GSTC boundary. Not limiting the maximum tangential wavenumber and/or not considering spatial dispersion in design can lead to disagreement between realized/measured and homogenized metasurfaces. The effects of not modeling spatial dispersion are explained in [6], where the authors employed conducting baffles to separate unit cells thereby eliminating non-local responses. The baffle approach was also adopted in [2], [3], [20], [24], to achieve agreement between the homogenized models and the realized metasurfaces. The solution presented in this paper is to use integral equations to model the metasurfaces. Integral equations inherently model non-locality through coupling matrices. Furthermore, they allow the true finite dimensions of metasurfaces to be modeled and can model curved surfaces easily. The beamforming metasurfaces considered here consist of a patterned metallic cladding (represented as a single electric impedance sheet) supported by a grounded dielectric substrate (see Fig. 1). The penetrable impedance sheet, the dielectric spacer, and ground plane are modeled using integral equations. It is shown that arbitrary far-field beams can be generated by simply patterning the upper metallic cladding layer of a commonly available microwave substrate, as depicted in Fig. 2. The cladding is patterned to control both the propagating and evanescent spectrum (surface waves) in order to beamform.

The surface waves needed to achieve beamforming are typically determined through optimization. Since the boundary condition used to model the beamforming metasurface relates the tangential electric field, $E^{tot}$, the sheet impedance, $\eta$, and the induced surface current density, $J$,

$$E^{tot} = \eta J, \tag{1}$$

the optimization variables can be chosen as any one of the three quantities. For example, in [2], [19], [25], the optimization variable is the induced surface current density, $J$. In [3], [4], [10], [11], [17], [21], the authors add explicitly defined surface wave fields, $E^{sw}$, and optimize their amplitudes. In [5], [7], [12], [13], [15], [16], [23] and in this work, the optimization is performed directly on the sheet impedances, $\eta$. In some cases, optimization considers both the impedances and the induced surface current simultaneously as in [14]. Table I summarizes recent scientific works on metasurface beamformer design, highlighting which quantity in (1) is chosen as the optimization variable.

There are many advantages to optimizing over the sheet impedances directly rather than the surface currents or surface-wave field amplitudes. Optimizing the impedances allows for a more direct connection to realized prototypes. Furthermore, optimizing the impedances does not require an a priori guess on which surface wave modes or which wavenumbers to include. The discretization simply determines the highest evanescent spectrum excited. In addition, the lossless property is enforced by making the impedances imaginary. However, two distinct disadvantages arise. Setting the discretization limits the range of realizable impedances. Second, as the metasurface grows in size, the number of unit cells can grow rapidly, and the optimization can become impractical. We have tackled both of these problems in this work. The maximum transverse wavenumber supported by the surface impedance is found based on its discretization. Using the dispersion relation for surface waves supported by an impedance sheet, the maximum impedance that can be realized is then related to this maximum transverse wavenumber. The maximum sheet impedance is used as a bound on the allowable sheet impedances used during optimization. In this way, the tangential wavenumber supported by the homogenized sheet in early phases of design can never exceed the maximum tangential wavenumber supported by the patterned metallic cladding of the metasurface. Consequently, a one-to-one correspondence is established between the homogenized design and its patterned metallic cladding realization. Second, we have accelerated our gradient descent optimization with a semi-analytic gradient calculation using the adjoint variable method. This allows the computational effort of the optimization to be independent of the number of unknown sheet impedances included in the optimization apart from the larger impedance matrix which must be inverted. These two innovations allow our models to more closely represent the realized structures, making the designs simple to realize. This in turn leads to designs which produce good agreement between measured or simulated and theoretical results.

The metasurface beamformers will be designed using a three-phase design strategy (see Fig. 2). This paper begins with an overview of the proposed three-phase design strategy in section II. The following three sections detail each design phase: phase 1 (*Direct solve*) is detailed in section III, phase 2 (*Optimization*) is detailed in section IV, and phase 3 (*Patterning*) is detailed in section V.

Three metasurface beamformer designs will be presented in section VI including a directly-fed compact metasurface antenna with nearly perfect aperture efficiency, a metasurface reflectarray designed for a scanned far-field beam with a -20 dB sidelobe level, and finally a conformal metasurface reflectarray. A study on the approximate losses and bandwidth of the beamforming metasurfaces is presented in section VII. The paper concludes with section VIII. Three appendices are provided. In the first appendix, derivations of the semi-analytic gradient of the adjoint variable method used in the gradient descent optimization of the metasurface beamformers is provided. In the second appendix, the adaptation of the method of moments (MoM) to the conformal metasurface design is presented. In the third appendix, geometrical parameters of the patterned metallic cladding are provided for the phase 3 realization of the directly-fed metasurface antenna. Finally, it should be noted that the presented examples in this paper include both reflective and directly-fed metasurfaces. Recently, the approach has been applied to the case of transmissive metasurfaces [23].

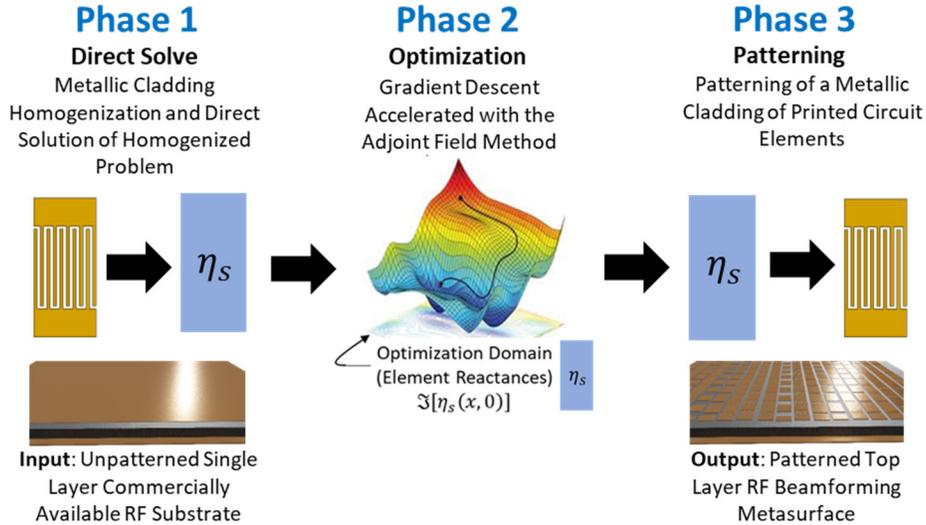

Fig. 2. Metasurface design strategy consisting of three phases: direct solve, optimization, and patterning phases.

## II. BEAMFORMING METASURFACE DESIGN OVERVIEW

The design of the metasurface is broken down into three design phases (see Fig. 2). In phase 1, the desired, total aperture field is defined. It may be defined directly from the known incident and desired scattered fields at the aperture, or it can be obtained from a desired far-field amplitude pattern using, for example, the Orchard-Elliott method. In either case, the patterned metallic cladding of the metasurface is homogenized and represented as a spatially variant array of homogenized impedance sheet elements linking the desired total field at the aperture to the induced surface currents through the impedance boundary condition (1). The impedance boundary condition is a limiting case of the GSTC's [1]. Then, a linearized electromagnetic inverse problem is solved to obtain the complex-valued impedances required to transform the incident field to the desired scattered field (see section III). The complex-valued impedances arise because the power density is not locally conserved at all points on the metasurface [13], even though global power conservation is enforced. Since the positive and negative resistances that arise are difficult or costly to manufacture, a purely reactive sheet is desired.

In phase 2, the complex-valued sheet is converted into a purely reactive sheet by way of gradient descent optimization accelerated by the adjoint variable method (see section IV). The reactances of the complex-valued sheet obtained in phase 1 serve as the critical initial optimization point required for convergence in local optimization-based techniques. The optimization varies the reactances until the normalized far-field amplitude scattered by the reactive sheet agrees with the normalized far-field amplitude scattered by the complex-valued sheet from phase 1. The optimization adds surface waves (evanescent fields) to the total field on the metasurface in order to satisfy a passive and lossless boundary condition.

In phase 3, the optimized reactive sheet impedances are realized as patterned metallic claddings. In phase 3, the optimized reactive sheet impedances are realized as patterned metallic claddings.

Note, in this work, an $e^{j\omega t}$ time convention is assumed and suppressed throughout. Bold face letters correspond to matrices (which may be matrices of dimension $N \times 1$) and arrow accents correspond to coordinate position vectors. The electromagnetics problem considered is 2-dimensional (with out-of-plane wavenumber $k_z=0$) with transverse electric (TE) polarization. Hence, all electric field and current density quantities are $\hat{z}$-directed, and all magnetic field quantities considered in the formulations are $\hat{x}$-directed.

## III. PHASE 1: DIRECT SOLVE

The geometry that is considered is illustrated in Fig. 1. A metasurface of width $w$ is placed above a grounded substrate with permittivity $\epsilon = \epsilon_r \epsilon_0 (1 - jtan\delta)$. The thickness of the substrate is $d$. The metasurface is divided into $N$ cells of width $w/N$ each. The metasurface consists of three layers (see Fig. 1). The first layer, denoted as layer 1, is the patterned metallic cladding, represented as an impedance sheet (metasurface), the second layer, denoted as layer 2, is the dielectric spacer, and the third layer, denoted as layer 3, is the PEC ground plane.

The task of design phase 1 is to design layer 1, the complex-valued inhomogeneous impedance sheet, to produce a desired field transformation. To that end, an electric field integral equation (EFIE) is written to model the response of the metasurface to the exciting field [26]. It includes the mutual coupling between the homogenized cells of the inhomogeneous impedance sheet, the polarization currents in the dielectric substate, and the surface currents on the ground plane [12], [13]. By applying the surface and volume equivalence principles, all the currents are assumed to radiate in free space. The integral equation has the form

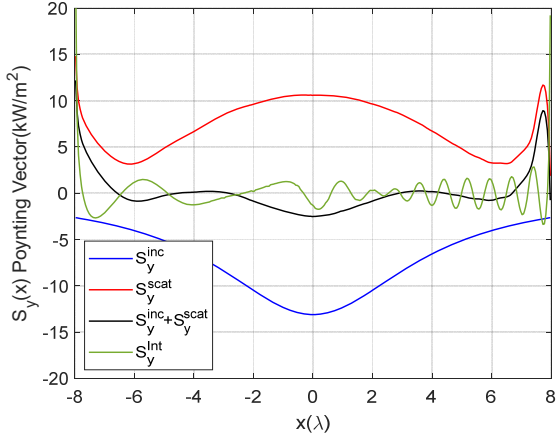

Figure 3. Normal component of Poynting vector at the metasurface plane associated with the incident, scattered, and interference between the incident and scattered fields for Metasurface Design B in section VI.B.

$$E_p^{inc}(\vec{\rho}_p) = \eta_p(\vec{\rho}_p) J_p(\vec{\rho}_p) + \sum_{q=1}^{3} j\omega\mu_0 \int G(\vec{\rho}_p, \vec{\rho}_q') J_q(\vec{\rho}_q') d\vec{\rho}_q', \quad (2)$$

where $\omega$ is the angular frequency of operation, $\mu_0$ is the magnetic permeability of free space, and $G$ is the free space Green's function for two-dimensional TE problems

$$G(\vec{\rho}, \vec{\rho}') = \frac{1}{4j} H_0^{(2)}(k_0 |\vec{\rho} - \vec{\rho}'|). \quad (3)$$

$H_0^{(2)}(\ )$ is the Hankel function of the second kind of order 0, and the indices $p$ and $q$ denote the layer numbers (see Fig. 1). When $p$ or $q$ is 1 or 3, $J_{p\ or\ q}$ is an electric surface current density and when $p$ or $q$ is 2, $J_{p\ or\ q}$ is the volumetric polarization current density. $\eta_p$ is the surface impedance of layer $p$, $\vec{\rho}_{p\ or\ q}$ is a position vector to a point in the $p$th or $q$th layer, and $E_p^{inc}$ is the incident field on layer $p$.

The incident field is that radiated by an infinite $\hat{z}$-directed electric line source placed $F$ meters above (or below) the metasurface along the $y$-axis. It is given by

$$E^{inc} = \frac{-I_o \eta_0 k_0}{4} H_0^{(2)}\left(k_0 \sqrt{x^2 + (y-F)^2}\right). \quad (4)$$

In (4), $I_o$ is the line current strength in Amperes, $\eta_0$ is the free space wave impedance, and $k_0$ is the free space wavenumber.

The desired scattered aperture field amplitude distribution is defined based on the desired far-field pattern. The amplitude level in V/m of the desired scattered aperture field is chosen to ensure global power conservation across the metasurface

$$\int_{-w/2}^{w/2} \frac{1}{2}\left[E_1^{sca} H_1^{sca*}\right] dx = -\int_{-w/2}^{w/2} \frac{1}{2}\left[E_1^{inc} H_1^{inc*}\right] dx = -P_y^{inc}, \quad (5)$$

where $P_y^{inc}$ is the power incident on the metasurface in Watts per unit length in $z$. From the desired scattered electric field, $E^{sca}$, the associated magnetic field is found by first computing the plane-wave spectrum of $E^{sca}$, namely $\tilde{E}_1^{sca}$, and then dividing the spectrum through by the TE wave impedance ($\eta_{TE} = \eta_0 k_0/k_y$) to obtain the spectrum of the magnetic field, $\tilde{H}_1^{sca}$.

Finally, an inverse Fourier transform is taken to obtain the magnetic field, $H_1^{sca}$. With $E_1^{sca}$ and $H_1^{sca}$ defined, the scattered power is found. The scattered electric field amplitude, $E_1^{sca}$, is scaled to satisfy (5). This ensures that the scattered field has the same total power as the incident field. This is a necessary condition for a metasurface to be passive and lossless.

With both $E_1^{inc}$ and $E_1^{sca}$ defined, (2) is solved for the unknown currents [12], [13] by making the substitution

$$\eta_1 J_1 = E_1^{inc} + E_1^{sca} = E_1^{tot}, \quad (6)$$

and noting that $\eta_2 = [j\omega\epsilon_0(\epsilon_r - 1)]^{-1}$ according to volume equivalence, and $\eta_3 = 0$ for the ground plane. The substitution in (6) linearizes the system, allowing for its direct solution using the MoM. Pulse basis functions and the Galerkin testing scheme are used in the MoM implementation (see the Appendix of [12] for the matrix element definitions). Applying the MoM, the linear system representation of (2) becomes [5], [12], [13]

$$\begin{bmatrix}[V_1]\\[V_2]\\[V_3]\end{bmatrix} = \begin{bmatrix}[\eta_1] & 0 & 0\\ 0 & [\eta_2] & 0\\ 0 & 0 & [\eta_3]\end{bmatrix}\begin{bmatrix}[I_1]\\[I_2]\\[I_3]\end{bmatrix} + \begin{bmatrix}[Z_{11}] & [Z_{12}] & [Z_{13}]\\[Z_{21}] & [Z_{22}] & [Z_{23}]\\[Z_{31}] & [Z_{32}] & [Z_{33}]\end{bmatrix}\begin{bmatrix}[I_1]\\[I_2]\\[I_3]\end{bmatrix}$$

$$V = \eta I + Z_c I = \{\eta + Z_c\} I = ZI$$

$$,(7)$$

Equation (7) can be solved by matrix inversion for the unknown induced currents. With the induced surface currents, (6) can be used again to obtain the sheet impedances $\eta_1$

$$\eta_1 = \frac{E_1^{tot}}{J_1}, \quad (8)$$

The obtained sheet impedances are, in general, complex-valued given that the desired total field may not satisfy a reactive impedance boundary condition. This is because the non-zero points of the normal component of the Poynting vector gives rise to local loss and gain [13], [27]. The Poynting vector contains three contributions as shown in Fig. 3. The figure shows the normal component of the Poynting vector associated with the incident field $S_y^{inc} = \frac{1}{2}Re[E_1^{inc} H_1^{inc*}]$, the scattered field $S_y^{scat} = \frac{1}{2}Re[E_1^{scat} H_1^{scat*}]$, and the interference between the incident and scattered fields $S_y^{int} = \frac{1}{2}Re[E_1^{inc} H_1^{scat*}] + \frac{1}{2}Re[E_1^{scat} H_1^{inc*}]$. The incident and scattered fields are taken from the example presented in section VI.B which transforms an incident cylindrical wave into a scanned and collimated beam with a prescribed sidelobe limit. Since $S_y^{inc} \neq S_y^{sca}$, the metasurface is reshaping the incident power density upon reflection, however, $P_y^{inc} = P_y^{sca}$ as indicated by (5). This key feature differentiates this work from the authors previous work in [5], [12], [13] where the incident power density was set equal to the scattered power density. Not restricting these power densities to be equal allows for far-field beamforming. The final term, the interference term, is associated with equal amounts of positive and negative power flux which integrates to zero. Consequently, the total integrated power is conserved since

$$\int_{-w/2}^{w/2} (S_y^{inc} + S_y^{sca} + S_y^{int}) dx = 0. \quad (9)$$

It is the non-zero local power density of $S_y^{tot} = S_y^{inc} + S_y^{sca} +$

$S_y^{int}$ which manifests itself locally as positive and negative resistances in the sheet impedances. Surface waves can be added to the scattered field to redistribute power transversally along the metasurface from places where local loss is needed to places where local gain is needed. According to (9), there should be just enough power to achieve a net local power density of zero at all points along the metasurface [9], [13], rendering the metasurface passive and lossless. Next, these surface waves are determined through optimization in phase 2.

## IV. PHASE 2: OPTIMIZATION

To introduce the surface waves through optimization, the following parameters of the optimizer are needed

    a) *Optimization variables and initial point*: The reactances of the complex-valued impedance sheets of phase 1 are retained to form the initial point.

    b) *Optimization domain boundaries*: The optimization domain limits are derived by limiting the capacitive sheet reactances such that the patterned metallic cladding supports the same surface waves as the homogenized impedance sheets.

    c) *Cost function*: The cost function is defined to be the Mean-Square-Error (MSE) between the far-fields of the metasurface resulting from phase 1 (taken to be the optimization goal) and that of phase 2.

    d) *Gradient computation*: The gradient is computed using the adjoint variable method.

The optimization parameters are detailed in the following subsections.

### A. Optimization variables and initial point

In gradient-based optimization methods, convergence strongly depends on obtaining a good initial point. A good initial point for the reactance of the impedance sheet can be obtained from the complex-valued impedance sheet from phase 1. This is accomplished by discarding the real parts of the complex-valued sheet impedances and retaining the reactances. These $N$ total unknown reactances are arranged in an $N$-dimensional space, where each reactance varies along an orthogonal axis. Since only reactances are retained in the optimization, the design is passive and lossless. This $N$-dimensional space defines the optimization domain.

### B. Optimization domain boundaries

Here, we derive bounds on the $N$ reactances that comprise the optimization domain. Since the optimization step introduces surface waves with tangential wavenumbers, $\beta_s$, outside the light cone, the surface-wave fields can vary across the patterned metallic cladding. The homogenized impedance sheet used to represent the patterned metallic cladding can theoretically support surface waves with large tangential wavenumbers (see Fig. 4b). However, the patterned metallic cladding has a tangential wavenumber cutoff which marks the onset of a stopband. Therefore, the maximum tangential wavenumber of the impedance sheet, $\beta_{s,max}$, must be limited such that the patterned metallic cladding supports the same surface waves as the homogenized impedance sheet. To find this limit, the frequency dispersion of the patterned metallic claddings can be investigated. Figure 4a shows the frequency dispersion of a unit cell of a patterned metallic cladding made from interdigitated capacitors (IDC), supported by a grounded Rogers 6010 substrate. The dispersion curves were obtained using the eigenmode solver of CST Microwave Studio (MWS). The substrate has a complex relative permittivity $\epsilon_r = 10.7(1 - j0.0023)$ and thickness $d = \lambda_0/23.62$ at 10 GHz. The unit cell dimensions are $0.05\lambda_0 \times 0.1\lambda_0$. The IDC has dimensions of $0.045\lambda_0 \times 0.1\lambda_0$. The dispersion for claddings with IDC teeth length corresponding to $l = 0.40, 0.59, 0.84, 1.27,$ and $2.16$ mm, is shown. The corresponding equivalent, homogenized sheet impedances are $\eta_s = -j50, -j40, -j30, -j20,$ and $-j10$ $\Omega$, respectively. The equivalent, homogenized sheet impedances were determined from Fig. 5a (see section V). As can be seen in Fig. 4a, at the operating frequency of 10 GHz, the patterned metallic cladding only supports surface waves up to an equivalent homogenized sheet impedance of approximately $-j20\ \Omega$. For sheet impedances less than $-j20\ \Omega$, the patterned metallic claddings support the same surface waves as their homogenized sheet equivalents. In Fig. 4b, it is shown that the dispersion curves of the patterned metallic claddings follow those of the equivalent homogenized impedance sheets. The dispersion curve of a homogenized impedance sheet is obtained using the Transverse Resonance Technique (TRT) [28]. Applying the TRT to an impedance sheet supported by a dielectric substrate with thickness $d$, and relative permittivity $\epsilon_r$, carrying a surface wave of the form

$$E^{sw} = Ae^{-j\beta_s x - \alpha y}, \qquad (10)$$

results in the dispersion relation

$$-j\frac{\sqrt{\beta_s^2 - k_0^2}}{\omega\mu_0} + Y_s - j\frac{\sqrt{\epsilon_r k_0^2 - \beta_s^2}}{\omega\mu_0}\cot\left(\sqrt{\epsilon_r k_0^2 - \beta_s^2}\, d\right) = 0. \qquad (11)$$

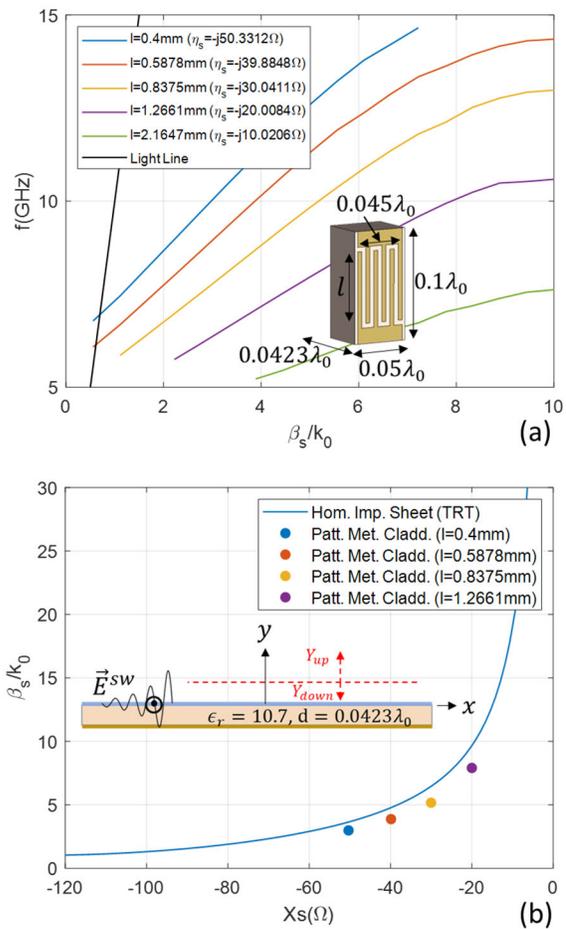

Figure 4. (a) Frequency dispersion curves for a patterned metallic cladding made from interdigitated capacitors and supported by a Rogers 6010 ($\epsilon_r$=10.7, $tan\delta$=0.0023, and $d$=0.00127 m) grounded dielectric substrate. Inset shows the unit cell used in the CST MWS Eigenmode simulations. (b) Tangential surface wavenumber normalized to the free space wavenumber versus sheet reactance. Inset shows the Rogers 6010 substrate used in the TRT. The filled dot colors correspond to the same-colored traces in (a). Note, TRT stands for Transverse Resonance Technique.

In (10) and (11), $\alpha$ is the attenuation constant satisfying the separation equation $\alpha^2 = \beta_s^2 - k_0^2$, and $Y_s = 1/\eta_s = 1/jX_s$ is the sheet admittance. The dispersion relation can be inverted to obtain a function relating the tangential wavenumber to the sheet impedance, $\beta_s(jX_s)$, at a given frequency $\omega_0$. This expression is plotted in Fig. 4b for $f$=10 GHz. The plot shows the tangential surface wavenumber normalized to the free space wavenumber for the homogenized impedance sheet. The eigenmode simulation results from the dispersion curves of the patterned metallic claddings of Fig. 4a are shown as colored dots in Fig. 4b. The colors of the dots correspond to the colors of the curves in Fig. 4a. The dots follow the TRT curve indicating that the patterned metallic claddings support the same surface waves as the homogenized sheets for reactances less than $-20\ \Omega$. Hence, all reactances between $-20\ \Omega$ and $0\ \Omega$ are excluded from the optimization to allow for realizable designs.

To exclude the reactances from the optimization domain, the optimization domain is bounded to the set

$$X_s \in [-1200\ \Omega, -20\ \Omega]. \qquad (12)$$

### C. Cost function

A surface is defined within the bounded optimization domain as $g(\mathbf{X_s}) = g(X_{s1}, X_{s2}, X_{s3}, \ldots, X_{sN})$ and represents the response of the metasurface as a function of its reactances. This surface is called the cost function. The cost function $g$ is designed such that its minimum represents the optimal solution. The cost function, $g(\mathbf{X_s})$, is defined as

$$g(\mathbf{X_s}) = \frac{1}{2}\Delta\mathbf{E}^\dagger \Delta\mathbf{E}, \qquad (13)$$

where † denotes Hermitian transpose and

$$\Delta\mathbf{E} = \frac{1}{\sqrt{M}}\left[20\log\left(\frac{|\mathbf{E}_{ff}^{sca,calc}|}{\|\mathbf{E}_{ff}^{sca,calc}\|_p}\right) - 20\log\left(\frac{|\mathbf{E}_{ff}^{sca,tar}|}{\|\mathbf{E}_{ff}^{sca,tar}\|_p}\right)\right], \qquad (14)$$

In (14), $\|\cdot\|_P$ is the $P$-norm, and $M$ is the number of far-field observation angles. $\mathbf{E}_{ff}^{sca,calc}$ and $\mathbf{E}_{ff}^{sca,tar}$ are the calculated and target scattered far-fields at the $M$ observation angles, respectively. They are calculated using

$$E_{ff}^{sca}(\rho,\phi) = -\frac{\eta_0 k_0}{4}\sqrt{\frac{2j}{\pi k_0 \rho}} e^{-jk_0\rho}\sum_{q=1}^{3}\iint_{S'} J_q(\vec{\rho}_q')e^{jk_0\vec{\rho}_q'\cdot\hat{\rho}}d\vec{\rho}_q'. \qquad (15)$$

In (15), $\rho$ and $\phi$ are the polar radial and angular coordinates, $\hat{\rho} = \cos\phi\hat{x} + \sin\phi\hat{y}$. The currents $J_q(\vec{\rho}_q')$ result from solving (7) for a given impedance vector. In particular, $E_{ff}^{sca,tar}(\rho,\phi)$ is found from the complex-valued impedance vector $\boldsymbol{\eta}$ from phase 1; and $E_{ff}^{sca,calc}(\rho,\phi)$ is found from the purely reactive impedance vector $j\mathbf{X_s}$ from phase 2. Note, $P$ (capitalized) refers to the p-value used in the $P$-norm whereas $p$ (lowercase) is used as a layer index. Also, the $P$-norm was chosen as it is a differentiable normalization function (for $P \neq 1$ or $\infty$). Note that the cost function must be differentiable in order to apply the adjoint variable method to the gradient calculation. For large enough $P$, the $P$-norm normalizes to the maximum value. In our work, we have found that $P = 250$ is sufficient. Cost function (44) minimizes the difference between the normalized far-field amplitude in dB scale scattered by the optimized purely reactive sheet (phase 2) and that of the complex-valued sheet of the direct solve design (phase 1). It should be mentioned that although using a normalized cost function does not require (5), enforcing global power conservation leads to a much better initial point and hence faster convergence.

### D. Gradient calculation

The gradient descent optimization used to optimize the sheets is accelerated by a semi-analytic gradient calculation using the adjoint variable method. The adjoint variable method formulation was adapted from that in [7], where it was successfully applied to metasurface optimization. Specifically, in [7], the adjoint variable method was applied to the circuit-based design of multi-input multi-output metastructures. Here, it is adapted to the design of metasurface beamformers. The detailed derivation of the gradient calculation can be found in appendix A and is summarized here. The adjoint variable method calculates an analytic gradient of (13) by forming its components as

$$\nabla g(\mathbf{X_s}) = \text{Re}\left\{-j\mathbf{I}_\lambda^\dagger \text{diag}(\mathbf{I})\right\}^T, \qquad (16)$$

where $\mathbf{I}$ represents the current density expansion coefficients obtained from the solution of the linear system in (7), $\text{diag}()$ refers to the creation of a diagonal matrix with the elements of the argument appearing along the diagonal, and $\mathbf{I}_\lambda$ is the solution to the adjoint problem

$$\mathbf{Z}^\dagger \mathbf{I}_\lambda = \mathbf{R}^\dagger \Delta \mathbf{E}^{adj}. \qquad (17)$$

In (17), $\mathbf{R}$ is an $M$-by-$N$ matrix formed by evaluating (15) at the $M$ observation angles in the far-field with the currents $J_q$ on all $N$ elements replaced by $1\angle 0°$ when pulse basis functions are used (see (A.17)). Also, $\mathbf{Z}$ is the method of moments impedance matrix in (7) and $\Delta\mathbf{E}^{adj}$ is the adjoint field defined in (A.17). The entire gradient is calculated by solving only 2 linear systems ((7) and (17)) at each iteration independent of the number of unknown sheet reactances $N$. The gradient in (16) is used in a classical gradient descent optimization method to optimize the impedance sheets. For details on the implementation of the gradient descent algorithm for metasurfaces, see [5].

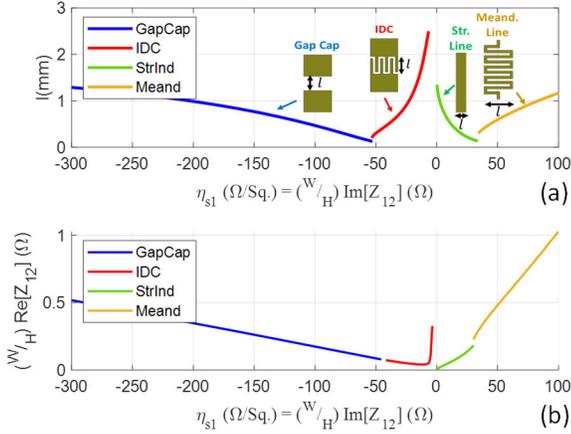

Fig. 5. Sheet impedance extraction of printed circuit elements in *periodic* environments. Each element is $W=0.9(\lambda_0/20)$ wide by $H=2(\lambda_0/20)$ tall. (a) Extracted sheet reactance versus geometrical parameter $l$. (b) Loss resistance as a function of sheet reactance.

The optimization results in purely reactive sheets which can be realized as a patterned metallic cladding using standard printed-circuit fabrication techniques. Now that phases 1 and 2 have been completed, the patterning phase, phase 3, can be applied.

## V. PHASE 3: PATTERNING

In this phase, the optimized, reactive metasurface from phase 2 is realized as a patterned metallic cladding of printed-circuit elements. Several printed-circuit elements were analyzed in order to extract the homogenized sheet impedance of the element [29]. The elements (shown in Fig. 5a) were assumed to have 18 $\mu$m thick copper (Cu). This thickness corresponds to a ½ ounce cladding that is commonly applied to microwave substrates. The elements were simulated in CST Microwave Studio using periodic boundaries and the unit cell geometry shown in Fig. 4a. The extracted reactances are shown in Fig. 5a, and the extracted resistances (representative of ohmic loss) as a function of the extracted reactances are shown in Fig. 5b. The resistances in Fig. 5b will be used in section VII.A to estimate the loss performance of the metasurface beamformers. Figure 5b is used to realize the sheet reactances resulting from the optimization applied during phase 2. The full metasurface beamformer is then simulated in COMSOL Multiphysics in a parallel plate waveguide for full-wave verification.

This completes the description of the three-phase design strategy. Three metasurface beamformer examples designed using this design strategy are presented next.

## VI. RESULTS

### A. Design A: Metasurface Antenna with Near Perfect Aperture Efficiency

The first example is a directly-fed metasurface antenna with nearly perfect aperture efficiency. This concept was first demonstrated in [30]. Here, we demonstrate the concept using a very thin substrate rather than a thick cavity and using a single electric impedance sheet. The aperture efficiency is the product of taper efficiency (a measure of how uniform the electric field amplitude is across the radiating aperture) and the radiation efficiency (a measure of the losses associated with the antenna). Thus, a metasurface antenna with near perfect aperture efficiency should exhibit a uniform amplitude aperture field and have near unity radiation efficiency. If the beam is also to be directed broadside, then the aperture field phase should be uniform (see Fig. 7). The metasurface that is considered is illustrated in Fig. 6. A metasurface of width $w=16\lambda_0$ ($f=10$ GHz) is placed above a conductor-backed Rogers 6010 substrate with $\epsilon_r$=10.7 and $tan\delta$=0.0023. The thickness of the substrate is $d$ =0.00127 m ($\lambda_0/23.62$). The metasurface is divided into $N$=320 cells of width $\lambda_0/20$ each. Each cell contains an impedance element of width $\Delta= 0.9\lambda_0/20$. An infinite line source feed is placed within the substrate at $F = -d/2$, as shown in Fig. 6. To define the desired scattered field amplitude, the incident power, $P_y^{inc}$, must first be calculated. To calculate $P_y^{inc}$, image theory is used to find the incident electric and magnetic fields, and hence the power density, at the metasurface plane. This power density is then integrated to obtain the power contained in the incident field

$$P_y^{inc} = \frac{1}{2} \int_{-w/2}^{w/2} \text{Re}\left[\left(E^{inc} + E_{image}^{inc}\right)\left(H^{inc} + H_{image}^{inc}\right)^*\right] dx, \quad (18)$$

where

$$E^{inc} = \frac{-I_o \eta_d k_d}{4} H_0^{(2)}\left(k_d\sqrt{x^2 + F^2}\right)$$

$$H^{inc} = -j\frac{I_o k_d}{4} \cos\left(\tan^{-1}\left(\frac{2x}{d}\right)\right) H_1^{(2)}\left(k_d\sqrt{x^2 + F^2}\right)$$

$$E_{image}^{inc} = \frac{I_o \eta_d k_d}{4} H_0^{(2)}\left(k_d\sqrt{x^2 + (F-d)^2}\right)$$

$$H_{image}^{inc} = j\frac{I_o k_d}{4} \cos\left(\tan^{-1}\left(\frac{2x}{3d}\right)\right) H_1^{(2)}\left(k_d\sqrt{x^2 + (F-d)^2}\right)$$

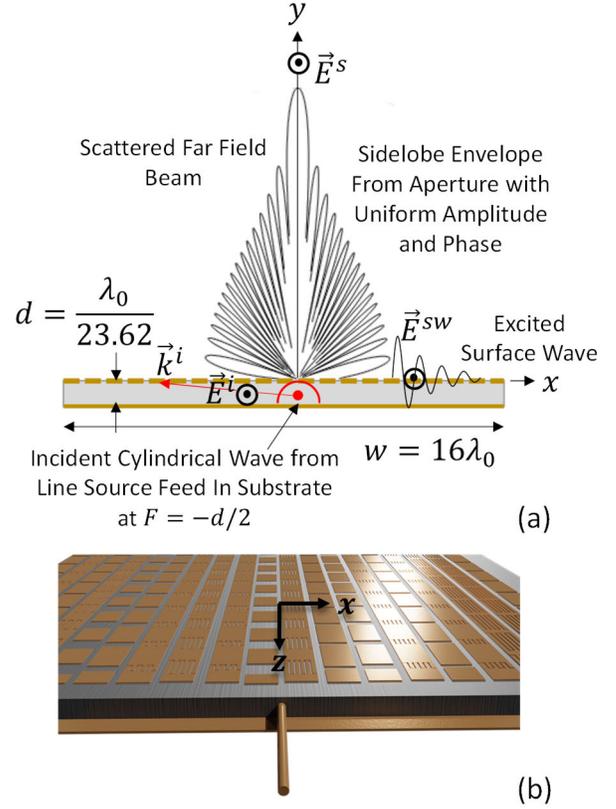

Fig. 6. Metasurface Design A): (a) Metasurface geometry. (b) Artistic rendition of a realized metasurface from phase 3. Note, the patterned metallic cladding is shown for illustrative purposes and does not correspond to an actual design.

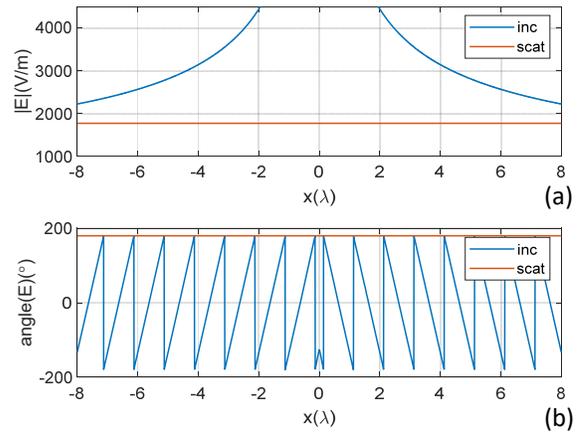

Fig. 7. Metasurface Design A: Incident and scattered field (a) amplitude and (b) phase at the metasurface plane.

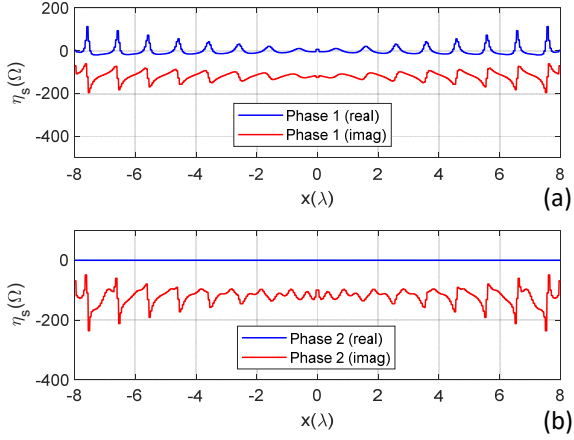

Fig. 8. Metasurface Design A: (a) Complex-valued sheet impedance of phase 1. (b) Optimized reactive sheet impedance of phase 2.

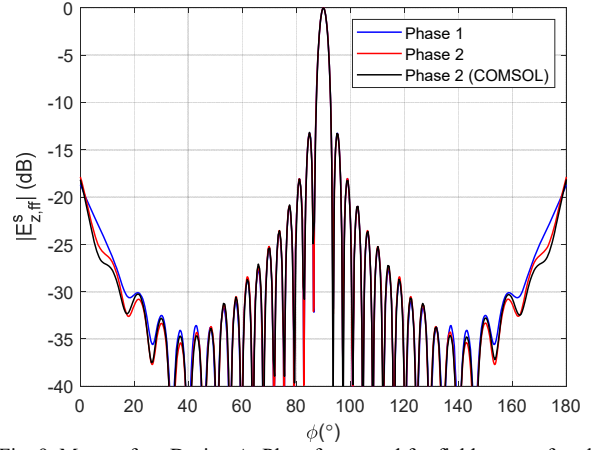

Fig. 9. Metasurface Design A: Plot of scattered far-field pattern for phase 1 and 2, and the COMSOL full-wave verification of the phase 2 design.

,(19)

and $k_d = k_0\sqrt{\epsilon_r}$ and $\eta_d = \eta_0/\sqrt{(\epsilon_r)}$. The desired amplitude (V/m) of the scattered aperture field is then found by solving (5), with $P_y^{inc}$ given by (18). Both the incident field due to this substrate-embedded line source and the uniform scattered aperture field amplitude and phase are shown in Fig. 7. In Fig. 8a, the complex-valued sheet impedances of phase 1 are shown. The real part shows both positive and negative resistances associated with equal amounts of positive and negative power flux in accordance with (5) and (9). The scattered far-field is shown in Fig. 9 and shows the expected *sinc*-like form with the

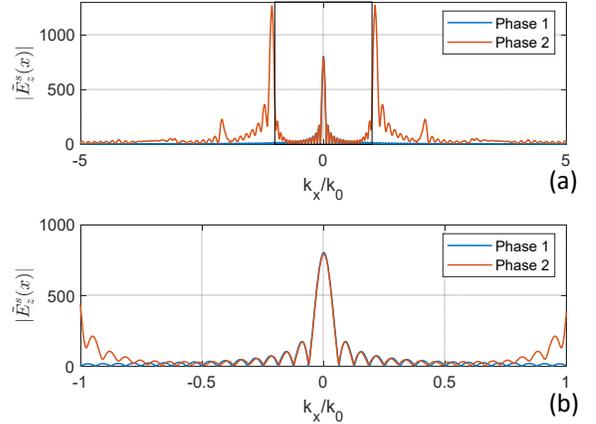

Fig. 10. Metasurface Design A: Scattered field amplitude spectrum at the plane of the metasurface for both phase 1 and phase 2. (a) Visible and invisible regions. (b) Visible region only.

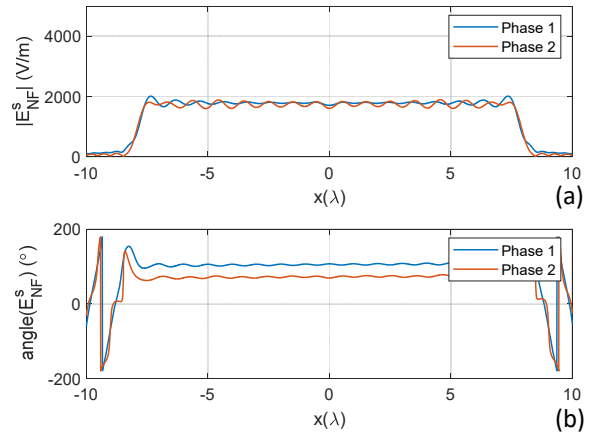

Fig. 11. Metasurface Design A: Far-fields back-projected to the metasurface plane. (a) Amplitude and (b) phase at the plane of the metasurface from both the phase 1 and phase 2 designs.

13.4 dB sidelobes characteristic of a uniform aperture. The radiation in the end-fire directions arises from edge diffraction and the feed power exiting the open substrate ends. These fields could be mitigated by including PEC via fences in the substrate at the ends forming a very thin cavity. The amplitude of the scattered electric field spectrum at the plane of the metasurface is shown in Fig. 10. The amplitude spectra are formed by calculating the FFT of the aperture fields along layer 1.

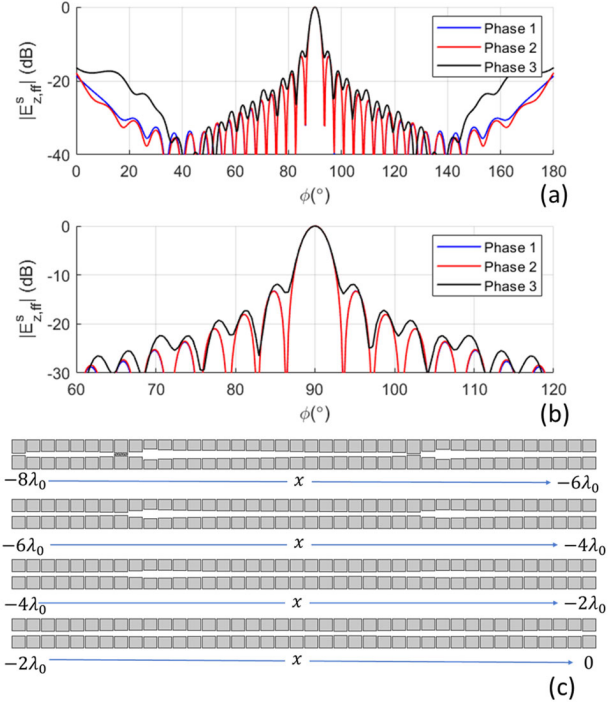

Fig. 12. Metasurface Design A: Plot of scattered far-field pattern for all three design strategy phases. (a) For $0° \leq \phi \leq 180°$. (b) For $60° \leq \phi \leq 120°$. (c) The left half of the simulated patterned metallic cladding, $-8\lambda_0 \leq x \leq 0$, of the symmetric metasurface. It is shown in four quarter sections containing 40 unit cells each.

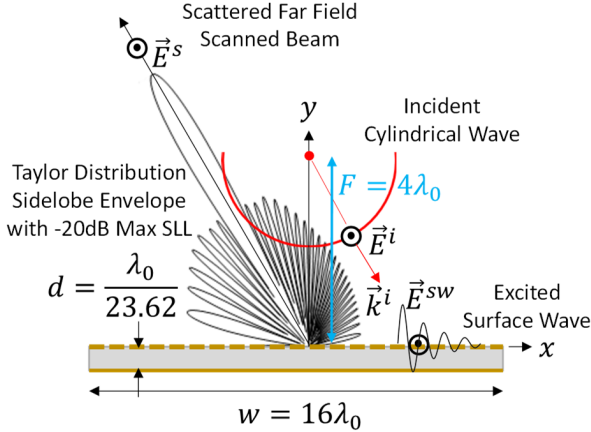

Fig. 13. Metasurface Design B: Metasurface geometry.

$$\widetilde{E}_1^{sca}(k_x) = \mathfrak{F}\left[E_1^{sca}\right] = \mathfrak{F}\left[\eta_1 J_1 - E_1^{inc}\right]. \qquad (20)$$

In (20), $\mathfrak{F}$ is the Fourier transform operator. By (6) and Fig. 7, this results in the phase 1 amplitude spectrum shown in Fig. 10. Both the invisible (Fig. 10a) and the visible (Fig. 10b) spectra are shown. It is observed that the spectrum of phase 1 does not include evanescent spectral content and hence there is no evidence of surface waves. Fig. 11 shows the far-field (see Fig. 9) back-projected to the metasurface plane. It shows that both the amplitude and phase are uniform, as stipulated.

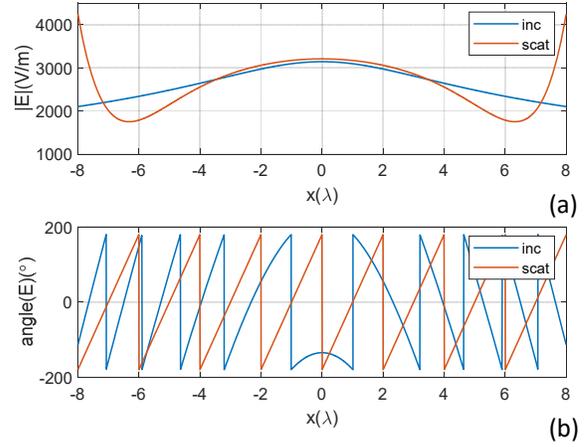

Fig. 14. Metasurface Design B: Incident and scattered field (a) amplitude and (b) phase at the metasurface plane.

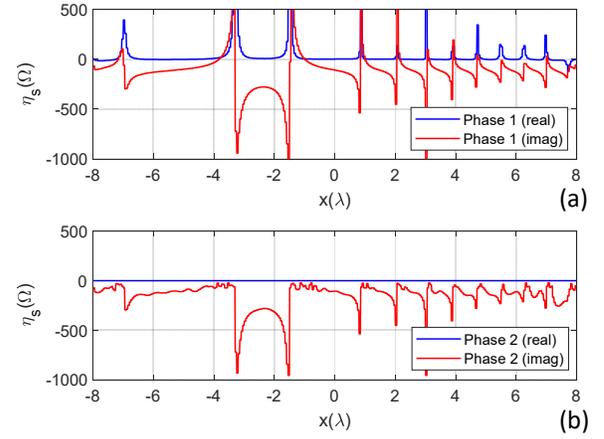

Fig. 15. Metasurface Design B: (a) Complex-valued sheet impedance of phase 1. (b) Optimized reactive sheet impedance of phase 2.

The purely reactive sheet impedances, resulting from phase 2, are shown in Fig. 8b. The perturbations to the reactance profile excite the surface waves needed to achieve passivity and losslessness. These surface waves can be seen in Fig. 10a where the evanescent fields introduced in phase 2 are shown in the invisible region. The visible region in Fig. 10b is identical to that of phase 1 apart from the visible region boundaries, where $k_x/k_0 = 1$. The discrepancy arises since $\mathfrak{I}[\eta_1 J_1 - E_1^{inc}]$ now also includes the surface waves. These surface waves also contribute to the edge diffracted fields. In Fig. 9, the far-fields are shown, and are seen to be indistinguishable from those of phase 1. Also shown in Fig. 9 is the COMSOL Multiphysics verification of the phase 2 design. The back-projected far-fields from both phases also agree well as indicated in Fig. 11.

As part of phase 3, the optimized reactances from phase 2, shown in Fig. 8b, were realized as printed circuit elements using Fig. 5a. The left half of the symmetric patterned metallic cladding is shown in Fig. 12c, and its geometrical details are provided in Table II in Appendix C. The patterned metallic clad-ding and grounded Rogers 6010 substrate were placed in a parallel plate waveguide and simulated in COMSOL Multiphysics. The results are shown in Fig. 12. The figure shows the far-field pattern comparison between all three design phases. The patterns agree well indicating that the three-phase design strategy is effective in designing metasurface beamformers that are realizable.

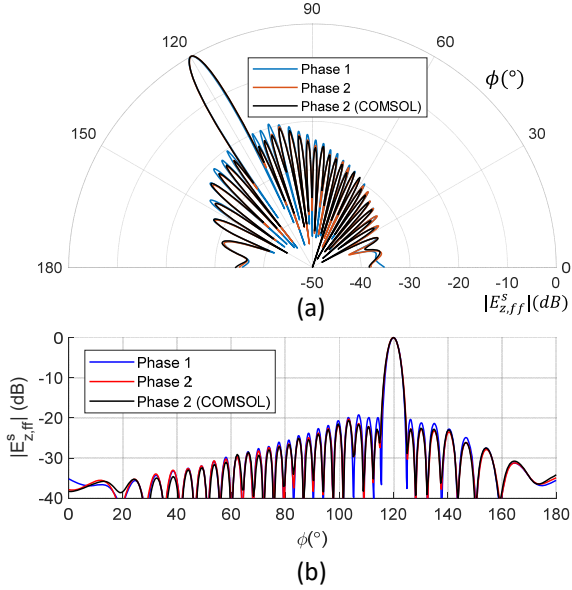

Fig. 16. Metasurface Design B: (a) Polar plot of scattered far-field pattern for both phase 1 and 2, and the COMSOL full-wave verification of the phase 2 design. (b) Rectangular plot for the same data.

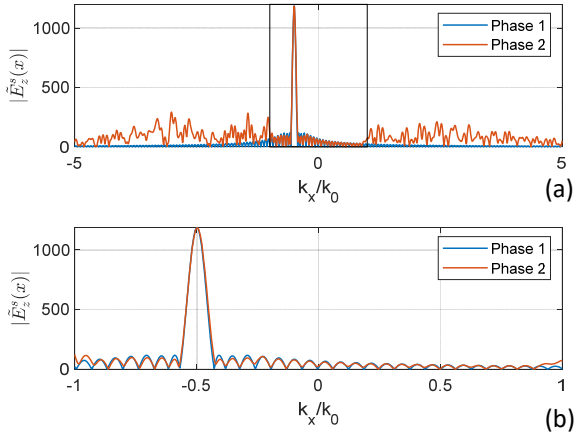

Fig. 17. Metasurface Design B: Scattered field amplitude spectrum at the plane of the metasurface for both phase 1 and 2. (a) Visible and invisible regions. (b) Visible region only.

In the next example, it is shown that the sidelobe envelope can be controlled in a metasurface reflectarray. This is, in general, not possible with traditional reflectarrays fed by non-tapered sources.

### B. Design B: Scanned Beam with -20dB Maximum Sidelobe Envelope

The next metasurface considered is illustrated in Fig. 13. A metasurface of width $w=16\lambda_0$ ($f=10$ GHz) is placed above the same conductor-backed Rogers 6010 substrate with $\epsilon_r=10.7$ and $\tan\delta=0.0023$. The thickness of the substrate is the same $d=0.00127$ m ($\lambda_0/23.62$). The metasurface is divided into $N=320$ cells of width $\lambda_0/20$ each. Each cell contains an impedance element of width $\Delta=0.9\lambda_0/20$. The metasurface is designed to transform a cylindrical wave radiated from a line source at $F=4\lambda_0$ to a collimated beam scanned to an angle of 30° from broadside, with a maximum sidelobe envelope of -20 dB. Hence, the amplitude of the aperture field is described by a Taylor distribution, and its aperture phase by a linear phase gradient. The incident power density can be obtained following from the expressions found in the appendix of [13]. After applying (5), the desired aperture field is shown in Fig. 14.

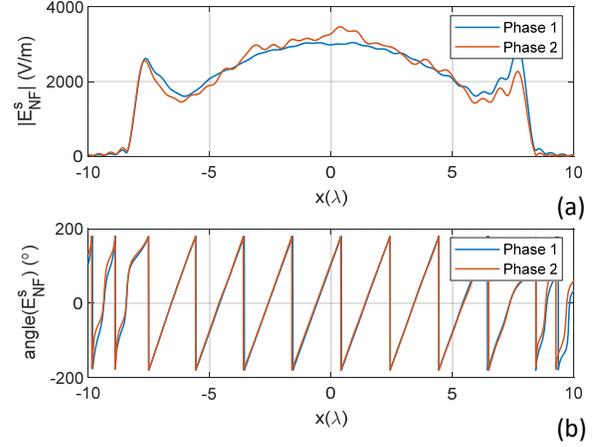

Fig. 18. Metasurface Design B: Far-fields back-projected to the metasurface plane. (a) Amplitude and (b) phase at the plane of the metasurface from both the phase 1 and phase 2 designs.

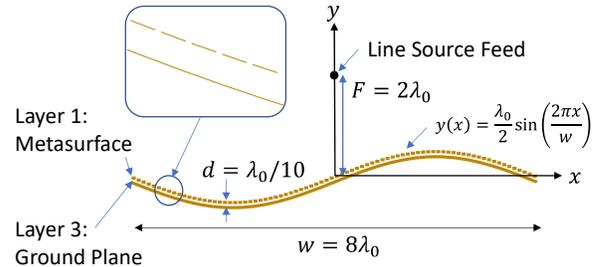

Fig. 19. Metasurface Design C: Conformal sinusoidal metasurface geometry. Note: In this case, layer 2 is removed as there is no dielectric spacer.

The results of phase 1 and phase 2 are shown in Figs. 15 through Fig. 18. The complex-valued sheet impedances of phase 1 and the purely reactive sheet impedances of phase 2 are compared in Fig. 15. Again, the perturbations in the reactances (see Fig. 15b) excite surface waves, represented by the evanescent spectrum in Fig. 17a. The visible spectrum of Fig. 17b, and the far-field of Fig. 16, show that the metasurface beamformer design of both phase 1 and phase 2 perform identically. The far-fields in Fig. 16 also include the COMSOL Multiphysics verification of the phase 2 design. The far-fields exhibit the desired scan angle and sidelobe envelope. In Fig. 18, the back-projected far-fields are shown, and excellent agreement is seen between the two phases. As can be observed from the amplitude of the back-projected field, the metasurface reshapes the power density as per Fig. 3. Hence, the surface waves not only allow for passivity and losslessness, but also reshape the incident power density upon reflection in order to beamform.

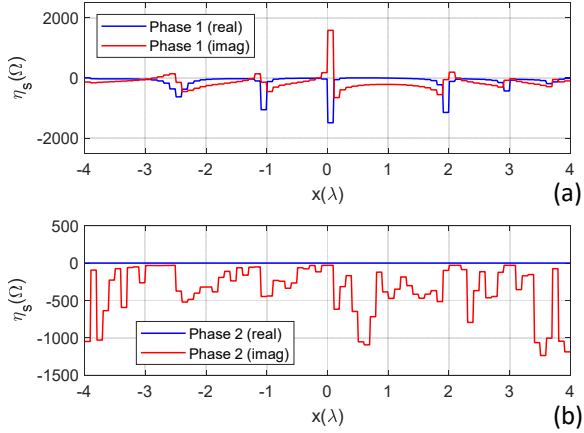

Fig. 20. Metasurface Design C: Complex-valued sheet impedance of phase 1. (b) Optimized reactive sheet impedance of phase 2.

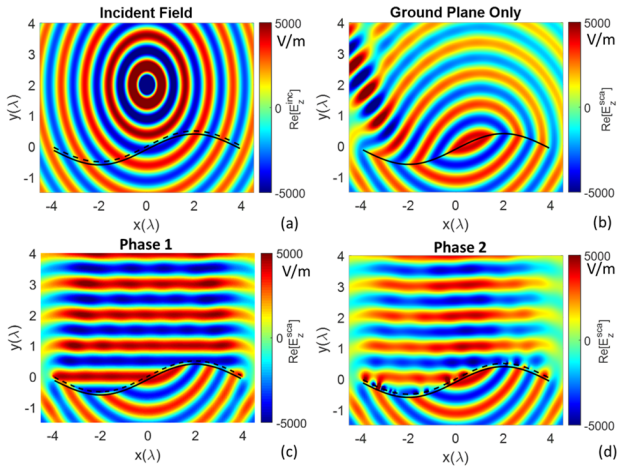

Fig. 21. Metasurface Design C: The real part of (a) the incident field, (b) the scattered field when only the ground plane is present, (c) the scattered near field for the phase 1 design, (d) the scattered near field for the phase 2 design.

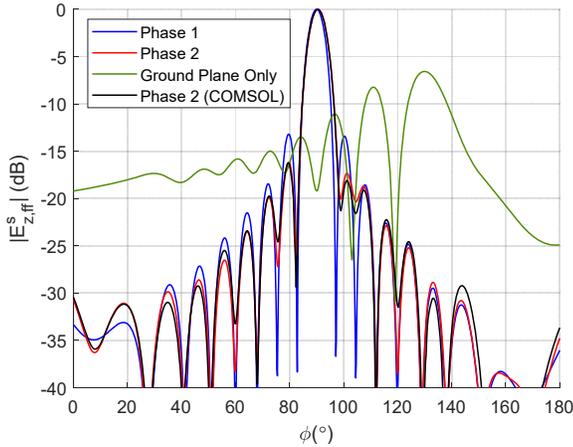

Fig. 22. Metasurface Design C: Far-field pattern comparison for phase 1 and 2, and the COMSOL full-wave verification of the phase 2 design.

In the next example, we show complete control of the scattered near and far-field from a metasurface that is conformal.

### C. Design C: Conformal Metasurface

The final example involves the conformal metasurface shown in Fig. 19. The metasurface is suspended in air at a distance of $\lambda_0/10$ above a conformal PEC ground plane. The $p$th layer of the conformal metasurface is described by the function

$$y_p(x) = \frac{\lambda_0}{2}\sin\left(\frac{2\pi x}{w}\right) - \delta_{p3}\frac{\lambda_0}{10}, \quad (21)$$

where $\delta_{pq}$ is the Kronecker delta function equal to unity when $p = q$ and zero otherwise. Note, $p = 1$ or 3 only as this metasurface has no dielectric spacer support. The curve in (21) is parameterized by the $x$-coordinate, and hence (2) can be specialized to the case of curvilinear metasurfaces as [31]–[35]

$$E_p^{inc}(x, y_p(x)) = \eta_p(x, y_p(x))J_p(x, y_p(x)) + $$
$$\frac{\eta_0 k_0}{4}\int_{-w/2}^{w/2}\left\{J_q(x', y_q(x'))H_0^{(2)}\left(k_0\sqrt{(x-x')^2 + (y_p(x)-y_q(x'))^2}\right)\right.$$
$$\left.\sqrt{1+\left(\frac{dy_q(x')}{dx'}\right)^2}dx'\right\}$$
$$(22)$$

The parameter space ($x$-axis) is divided up into $N=80$ cells of equal width that are each $\lambda_0/10$ wide, hence $w = 8\lambda_0$. For the metasurface layer (layer 1), each cell contains an impedance sheet of width $\Delta = 0.9\lambda_0/10$. For the ground plane (layer 3), the cells are completely filled with PEC. The cells project conformally onto the metasurface using the parameterization in (21). Hence, for example, the $l$th impedance sheet on layer 1 has an arc length of $S_\Delta(x) = \int_{x_l-\Delta/2}^{x_l+\Delta/2}\sqrt{1+(dy_1/dx)^2}\,dx$. The metasurface is designed to generate a uniform amplitude and phase aperture field. The metasurface is fed by a line source placed $F = 2\lambda_0$ above its center, as depicted in Fig. 19. Using the formulation in Appendix B, (22) is solved for the unknown induced current densities as part of design phases 1 and 2. The results of phases 1 and 2 are shown in Fig. 20 through Fig. 22. The impedances shown in Fig. 20 are plotted against the $x$-coordinate of each of the elements. In Fig. 21 the real part of the incident near field (21a), the real part of the scattered near field when only the ground plane is present (21b), the real part of the scattered near field for the complex-valued metasurface from phase 1 (21c), and the real part of the scattered near field for the optimized reactive metasurface from phase 2 (21d) are shown. The figure shows planar wavefronts scattered from the conformal metasurface. In Fig. 22, the far-fields for phases 1 and 2 are shown, as well as the far-fields of the case of the ground plane only and the COMSOL Multiphysics verification of the phase 2 design. The far-fields show excellent agreement between the phase 1 and phase 2 metasurface designs. Thus, the design approach presented in this paper allows for the complete control of the scattered radiative near and far-fields from conformal surfaces of any shape. This is critical for next-generation cloaks, electromagnetic illusions, and camouflage.

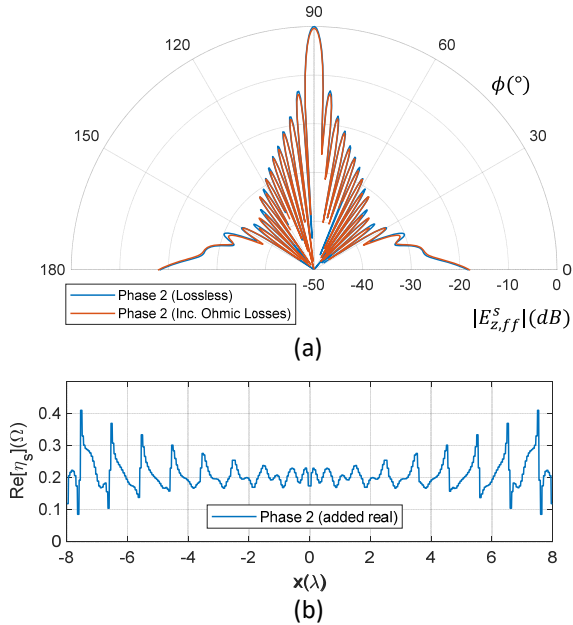

Fig. 23. Metasurface Design A: (a) Far-field comparison between the phase 2 design with and without added ohmic (dielectric and metal losses). (b) The added Cu loss implemented as a real part of the homogenized sheet impedances.

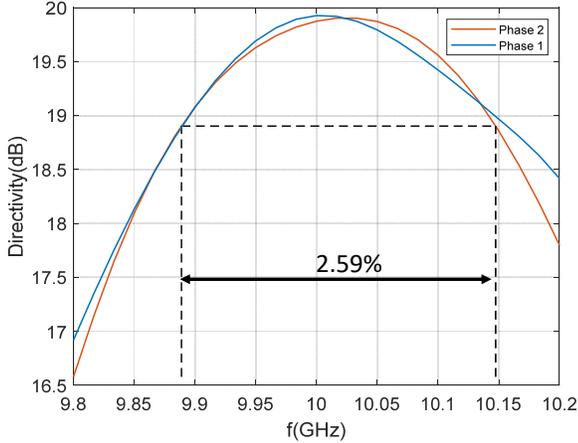

Fig. 24. Metasurface Design A: Directivity versus frequency for both the phase 1 and phase 2 designs. The 1dB directivity bandwidth of the phase 2 design is indicated in the figure as 2.59%.

Since the metasurface is not planar, locally periodic sheet impedance extraction techniques, which rely on periodic field expansions, cannot be used to realize the metasurface. In this case, the aperiodic unit cell design technique presented in [36], [37] may be used to realize the metasurfaces as a conformal patterned metallic cladding.

## VII. ESTIMATION OF LOSS AND BANDWIDTH

### A. Loss

In order to characterize the performance of metasurfaces realized with practical materials, there are two sources of loss that need to be considered: (1) dielectric losses and (2) metal losses. The dielectric losses are modeled with a complex permittivity in the volumetric impedance $\eta_2$ of (7). Figure 5 is used to estimate the metallic loss (shown in Fig. 23b) associated with each homogenized sheet element of the optimized reactive metasurface of section VI.A (shown in Fig. 8b). The loss was modeled as a real part added to the optimized sheet impedances of phase 2. Two COMSOL Multiphysics simulations were performed, one of a completely lossless antenna, and another with both dielectric and ohmic losses included. The resultant far-field patterns are compared in Fig. 23a. The radiation efficiency was calculated to be $e_r = 0.9792$. It was found by forming the ratio of $P_{rad,lossy}$ to $P_{rad,lossless}$. It shows that the metasurface beamformers are not sensitive to ohmic losses and achieve near-perfect aperture efficiency. This is partly due to the exclusion of inductive reactances which carry high current densities on thin conductive traces (see Fig. 5b).

### B. Bandwidth

The bandwidth of the metasurface beamformer example is calculated by performing frequency sweep simulations of the metasurface considered in section VI.A using COMSOL Multiphysics. The optimized reactances in Fig. 8b were scaled at each frequency in the sweep according to

$$\text{Im}[\eta_s]_f = \begin{cases} \text{Im}[\eta_s]_{10GHz}\left(\dfrac{10GHz}{f}\right), & \text{if } \text{Im}[\eta_s]_{10GHz} < 0 \\ \text{Im}[\eta_s]_{10GHz}\left(\dfrac{f}{10GHz}\right), & \text{if } \text{Im}[\eta_s]_{10GHz} > 0 \end{cases} \quad (23)$$

The directivity of the far-field radiation pattern was calculated at each frequency. The same sweeps were performed for the complex-valued sheet design from phase 1 of Fig. 8a. In those sweeps, the reactances were scaled according to (23) while the resistances were kept constant. A plot of the far-field pattern directivity versus frequency for both cases is shown in Fig. 24. The metasurface beamformer example shows a 2.59% 1 dB directivity bandwidth. For comparison, the 1 dB directivity bandwidth for the phase 1 design is 2.65%.

## VIII. CONCLUSION

An approach for the design of beamforming metasurfaces was presented. The metasurfaces considered consist of a spatially-varying array of homogenized impedance sheets supported by a finite, truncated grounded dielectric substrate. They are modeled using integral equations. The design approach consists of three phases, an initial *direct solve* phase that solves for a complex-valued impedance sheet, a subsequent *optimization* phase to remove real parts of the sheet impedances obtained as a result of step 1, and a final *patterning* phase. The three-phase design approach results in passive and lossless metasurfaces which can easily be realized with patterned metallic claddings.

The key to realizing passive and lossless metasurfaces capable of arbitrary field transformation is the optimization phase. In this phase, surface waves are introduced that transfer power transversally from places where loss is needed to places where gain is needed. This leads to a passive design which reshapes the power density upon reflection. The optimizations of the sheet reactances were performed using gradient descent accelerated by supplying a semi-analytic gradient obtained using the adjoint variable method. Examples of both planar and conformal beamforming metasurface reflectarrays were reported. In addition, a directly-fed metasurface antenna with near-perfect aperture efficiency was reported.

Finally, the bandwidth and sensitivity to losses was investigated. The range of allowable reactances were limited in the optimization, avoiding printed circuit elements in the patterned metallic cladding which were associated with high metallic losses. This resulted in beamformers less sensitive to ohmic losses. It was also found that the metasurface beamformers exhibited bandwidths exceeding 2.59%, making them viable candidates for communications links or for electromagnetic illusions, and camouflaging. Future work includes extending the design concepts to 3D designs and increasing the bandwidth of the beamformers.

## A. APPENDIX: ADJOINT VARIABLE OPTIMIZATION EXPRESSIONS

The adjoint variable optimization expressions are derived in this appendix. The optimization variables are the reactances of the metasurface elements

$$\bm{X}_s = [X_{s1}, X_{s2}, \ldots, X_{sN}]^T. \qquad (A.1)$$

The cost function is defined in (13) and is repeated here as

$$g(\bm{X}_s) = \frac{1}{2}\Delta \bm{E}^\dagger \Delta \bm{E}. \qquad (A.2)$$

If the cost function is to compare un-normalized complex-valued far electric fields, then

$$\Delta \bm{E} = \begin{pmatrix} E_{ff}^{sca,calc}(r_1) \\ \vdots \\ E_{ff}^{sca,calc}(r_m) \\ \vdots \\ E_{ff}^{sca,calc}(r_M) \end{pmatrix} - \begin{pmatrix} E_{ff}^{sca,tar}(r_1) \\ \vdots \\ E_{ff}^{sca,tar}(r_m) \\ \vdots \\ E_{ff}^{sca,tar}(r_M) \end{pmatrix}, \qquad (A.3)$$

$$= \bm{E}_{ff}^{sca,calc}(\vec{r}) - \bm{E}_{ff}^{sca,tar}(\vec{r})$$

where $E_{ff}^{sca}$ is found from (15) and $r_m = (\rho_m, \phi_m)$. The derivative of the cost function with respect to the $k^{th}$ optimization variable is

$$\frac{\partial g(\bm{X}_s)}{\partial X_{sk}} = \frac{\partial}{\partial X_{sk}}\left\{\frac{1}{2}\Delta \bm{E}^\dagger \Delta \bm{E}\right\} = \frac{1}{2}\left\{\Delta \bm{E}^\dagger \frac{\partial \Delta \bm{E}}{\partial X_{sk}} + \frac{\partial \Delta \bm{E}^\dagger}{\partial X_{sk}}\Delta \bm{E}\right\}$$
$$(A.4)$$

By the identity $a^*b + ab^* = 2Re[a^*b]$, (A.4) becomes

$$\frac{\partial g(\bm{X}_s)}{\partial X_{sk}} = \text{Re}\left\{\Delta \bm{E}^\dagger \frac{\partial \Delta \bm{E}}{\partial X_{sk}}\right\}$$
$$= \text{Re}\left\{\Delta \bm{E}^\dagger \frac{\partial}{\partial X_{sk}}\left[\bm{E}_{ff}^{sca,calc} - \bm{E}_{ff}^{sca,tar}\right]\right\} \quad .(A.5)$$
$$= \text{Re}\left\{\Delta \bm{E}^\dagger \frac{\partial \bm{E}_{ff}^{sca,calc}}{\partial X_{sk}}\right\}$$

Writing (15) as an operator $T[\bm{J}]$ which transforms a row vector of current density, $\bm{J} = [J_1, J_2, J_3]$, to a column vector of electric far-field values at the points $(r_1, \ldots, r_m, \ldots, r_M)^T$, (A.5) becomes

$$\frac{\partial g(\bm{X}_s)}{\partial X_{sk}} = \text{Re}\left\{\Delta \bm{E}^\dagger \frac{\partial \bm{E}_{ff}^{sca,calc}}{\partial X_{sk}}\right\}$$
$$= \text{Re}\left\{\Delta \bm{E}^\dagger \frac{\partial T(\bm{J})}{\partial X_{sk}}\right\} = \text{Re}\left\{\Delta \bm{E}^\dagger T\left[\frac{\partial \bm{J}}{\partial X_{sk}}\right]\right\}, \qquad (A.6)$$

where the last equality follows from the linearity of $T$. Next, we formulate an expression for $\partial \bm{J}/\partial X_{sk}$. Using the solution to the linear system in (7), (A.6) becomes

$$\frac{\partial \bm{J}}{\partial X_{sk}} = \frac{\partial}{\partial X_{sk}}\{\bm{BI}\}, \qquad (A.7)$$

where $\bm{B} = [b_1|b_2|\ldots|b_N]$, and $b$ are the basis functions (in our cases simple pulse basis functions). Note, the '|' symbol means to juxtapose the basis functions, $b$, into a row vector, and hence, the product of $\bm{BI}$ is an element wise product. Since $\bm{B}$ is not a function of $X_{sk}$, (A.7) becomes

$$\frac{\partial \bm{J}}{\partial X_{sk}} = \bm{B}\frac{\partial \bm{I}}{\partial X_{sk}} = \bm{B}\frac{\partial}{\partial X_{sk}}\{\bm{Z}^{-1}\bm{V}\}$$
$$= \bm{B}\left\{\frac{\partial \bm{Z}^{-1}}{\partial X_{sk}}\bm{V} + \bm{Z}^{-1}\frac{\partial \bm{V}}{\partial X_{sk}}\right\}. \qquad (A.8)$$

By the rule for differentiation of the inverse of a matrix [38] and since $\bm{V}$ is not a function of $X_{sk}$

$$\frac{\partial \bm{J}}{\partial X_{sk}} = \bm{B}\left\{-\bm{Z}^{-1}\frac{\partial \bm{Z}}{\partial X_{sk}}\bm{Z}^{-1}\bm{V}\right\} = \bm{B}\left\{-\bm{Z}^{-1}\frac{\partial \bm{Z}}{\partial X_{sk}}\bm{I}\right\}. \quad (A.9)$$

Since $\bm{Z} = \bm{Z}_c + \bm{\eta} = \bm{Z}_c + j\bm{X}$, (A.9) becomes

$$\frac{\partial \bm{J}}{\partial X_{sk}} = \bm{B}\left\{-\bm{Z}^{-1}\frac{\partial}{\partial X_{sk}}[\bm{Z}_c + j\bm{X}]\bm{I}\right\}$$

$$= \bm{B}\left\{-j\bm{Z}^{-1}\begin{bmatrix} 0 & \cdots & 0 \\ & \ddots & \\ \vdots & 1 & \vdots \\ & & \ddots \\ 0 & \cdots & 0 \end{bmatrix}\bm{I}\right\} = \bm{B}\left\{-j\bm{Z}^{-1}\begin{bmatrix} 0 \\ \vdots \\ I_k \\ \vdots \\ 0 \end{bmatrix}\right\}$$
,(A.10)

where we have explicitly factored out the $j$ from the reactive impedances in the diagonal matrix $\bm{\eta}$. Substituting back into (A.6),

$$\frac{\partial g(\bm{X}_s)}{\partial X_{sk}} = \text{Re}\left\{-j\Delta \bm{E}^\dagger T\left[\bm{BZ}^{-1}\begin{bmatrix} 0 \\ \vdots \\ I_k \\ \vdots \\ 0 \end{bmatrix}\right]\right\}. \qquad (A.11)$$

The result (A.11) is for the $k^{th}$ component of the gradient. For the full gradient, we form a matrix with (A.11) as the $k^{th}$ column

$$\nabla g(\bm{X}_s)^T = \text{Re}\left\{-j\Delta \bm{E}^\dagger T\left[\bm{BZ}^{-1}\text{diag}(\bm{I})\right]\right\}. \qquad (A.12)$$

Since $T$ in (A.12) is a linear operator, its action can be represented with a matrix. By combining $T$ and $\bm{B}$, a matrix $\bm{R}$ can be formulated with the $m^{th}$ observation point appearing along the

$m^{th}$ row and the $n^{th}$ basis function appearing on the $n^{th}$ column as (and hence $R$ is of size $M \times N$)

$$R = T[B]. \quad (A.13)$$

The matrix $R$ is defined such that $E_{ff}^{sca} = RI$. Equation (A.13) allows (A.12) to be written as

$$\nabla g(X_s)^T = \text{Re}\{-j\Delta E^\dagger R Z^{-1} diag(I)\}. \quad (A.14)$$

The adjoint equation is therefore

$$\Delta E^\dagger R Z^{-1} = I_\lambda^\dagger$$
$$Z^\dagger I_\lambda = R^\dagger \Delta E \quad (A.15)$$

The quantity $\Delta E$ in (A.15) is called the adjoint field $\Delta E^{adj}$. The linear system of the adjoint problem in (A.15) can be solved leading to the final form for the gradient of the cost function

$$\nabla g(X_s)^T = \text{Re}\{-jI_\lambda^\dagger diag(I)\}. \quad (A.16)$$

Only the first $N$ components of the vector in (A.16) are kept since the remaining are zero given that the optimization variables only exist along the first layer. Thus, to calculate the full gradient, the linear system in (7) is solved at each new optimization point yielding $Z$ and $I$ (the MoM impedance matrix and current vector of the forward problem solution). Subsequently, $\Delta E$ is found from this result and (A.3). Finally the adjoint equation is solved for $I_\lambda^\dagger$ yielding the full gradient at the optimization point following from (A.16). Note, $R$ is calculated from (A.13) and can be precalculated only once for the entire optimization. In this way, the entire gradient is found by solving only the forward linear system (7) and the adjoint linear system of (A.15), independent of the length of (A.1) (the number of unknowns in the optimization).

When the cost function is formulated to compare the normalized far-field amplitude in dB scale as in (14) rather than the complex-valued far field as in (A.3), the derivation is the same except the adjoint field $\Delta E$ in the right hand side of (A.15) becomes

$$\Delta E^{adj} = \frac{20}{\ln[10]} \frac{1}{\|E_{ff}^{sca,calc}\|_P} \left(\Delta E \oslash |E_{ff}^{sca,calc}|_n - \left(\sum_m \Delta E_m\right)\left(\sum_m |E_{ff}^{sca,calc}|_{n,m}^P\right)^{\frac{1-P}{P}} \left(|E_{ff}^{sca,calc}|_n^{P-1}\right)\right) \odot, \quad (A.17)$$

$$\left(E_{ff}^{sca,calc}(\rho,\phi) \oslash |E_{ff}^{sca,calc}|\right)$$

where a subscript $n$ indicates a quantity normalized to its own maximum, $|E_{ff}^{sca,calc}|_n^{P-1}$ is the Hadamard (element-wise) product of $|E_{ff}^{sca,calc}|_n$ with itself $P-1$ times. Note, $\odot$ and $\oslash$ are the Hadamard (element-wise) multiplication and division operators, respectively. In the work presented in this paper, a value of $P=250$ was chosen.

## B. APPENDIX: CONFORMAL METHOD OF MOMENTS

The unknown induced current density in (22) is expanded using pulse basis functions similar to the planar case. However, the pulse basis functions in the curvilinear case, $s_n$, are defined as [35], [39]

$$s_n(x, y_p(x)) = \begin{cases} \frac{1}{\sqrt{1 + \left(\frac{dy_p(x)}{dx}\right)^2}}, & x_n - \Delta x \le x \le x_n + \Delta x \\ 0, & \text{otherwise} \end{cases} \quad (B.1)$$

The quantity $\Delta x$ denotes the basis function length along the $x$-axis (the parameter space). We also define the normalization function, which makes the results of integrations over the testing functions equal to average field values (see (B.4))

$$S_n(x, y_p(x)) = \int_{x_n - \Delta x}^{x_n + \Delta x} s_n(x, y_p(x)) dx. \quad (B.2)$$

The unknown current density is then expanded as

$$J_p(x, y_p(x)) = \sum_{n=1}^N I_p^n s_n(x, y_p(x)) = \sum_{n=1}^N \frac{I_p^n}{\sqrt{1 + \left(\frac{dy_p(x)}{dx}\right)^2}}. \quad (B.3)$$

The reason for this definition is that it simplifies the integrations required to compute the MoM voltage vector and impedance matrix elements since the Jacobian in (22) cancels out with the denominator of (B.3). Hence, substitution of (B.3) into (22) and using the Galerkin MoM formulation leads to the following matrix element definitions

$$[V_p^m] = S_m \int_{x_m - \Delta x}^{x_m + \Delta x} E_p^{inc}(x, y_p(x)) dx$$

$$[\eta_1^{mn}] = \delta_{mn} \eta_s^n S_m \int_{x_m - \Delta x}^{x_m + \Delta x} \frac{1}{\sqrt{1 + \left(\frac{dy_p(x)}{dx}\right)^2}} dx$$

$$[Z_{pq}^{mn}] = S_m \frac{\eta_0 k_0}{4} \int_{x_m - \Delta x}^{x_m - \Delta x} \int_{x_n - \Delta x}^{x_n - \Delta x} H_0^{(2)}\left(k_0 \sqrt{(x-x')^2 + (y_p(x) - y_q(x'))^2}\right) dx' dx$$

$$.(B.4)$$

The self-impedance terms are found as

$$[Z_{pp}^{nn}] = S_n \frac{\eta_0 k_0 (\Delta x)^2}{4} + jS_n \frac{\eta_0 k_0 (\Delta x)^2}{4\pi}[3 + \ln(4) - 2\ln(\gamma k_0 \Delta x)]$$

$$-S_n \frac{j\eta_0 k_0}{4\pi} \int_{x_m - \Delta x}^{x_n - \Delta x} \int_{x_n - \Delta x}^{x_n - \Delta x} \ln\left[1 + \left(\frac{\pi \lambda_0}{w}\cos\left(\frac{\pi}{w}(x+x')\right)\right)^2\right] dx' dx$$

$$,(B.5)$$

where $\gamma = 1.781$ is Euler's constant. The remaining integration in (B.5) is non-singular and can be done numerically. The vectors and matrices of (B.4) and (B.5) can be inserted into (7) (with all $p$ or $q = 2$ rows/columns removed from (7)) and solved for the unknown induced currents. Owing to (B.3), (8) becomes

$$\eta_s^n = \frac{E_1^{tot}(x_n, y_1(x_n))}{J_1(x_n, y_1(x_n))} = \frac{E_1^{tot}(x_n, y_1(x_n))\sqrt{1 + \left(\frac{\lambda_0 \pi}{w}\cos\left(\frac{2\pi x_n}{w}\right)\right)^2}}{I_1^n}$$

$$. \quad (B.6)$$

Finally, the far-fields are calculated by integrating the currents along the curvilinear surface. Hence, (15) becomes

$$E_{ff}^{sca}(\rho,\phi) = -\frac{\eta_0 k_0}{4}\sqrt{\frac{2j}{\pi k_0 \rho}} e^{-jk_0\rho}$$

$$\left\{ \int_{-w/2}^{w/2} J_1(x, y_1(x)) e^{jk_0(x\cos\phi + y_1(x)\sin\phi)} \sqrt{1 + \left(\frac{dy_1(x)}{dx}\right)^2} dx + \right.$$

$$\left. \int_{-w/2}^{w/2} J_3(x, y_3(x)) e^{jk_0(x\cos\phi + y_3(x)\sin\phi)} \sqrt{1 + \left(\frac{dy_3(x)}{dx}\right)^2} dx \right\}$$

.(B.7)

Equation (A.13) is similarly modified accordingly.

## C. APPENDIX: DIMENSIONS OF THE PRINTED CIRCUIT ELEMENTS

This appendix lists the dimensions of the printed circuit elements comprising the patterned metallic cladding of Metasurface Design A in Fig. 12c. The columns labeled $n$ list the indices of the element starting from the leftmost end of the metasurface at $x = -8\lambda_0$ and progressing toward the origin. The centroid of each element can then be found from the index as $x_{n,centroid} = -8\lambda_0 + (n - 1/2)\lambda_0/20$. The columns labeled $l$ list the geometrical parameter $l$ (in units of mm) of the interdigitated capacitors (IDC) and gap capacitors (GapCap) that make up the patterned metallic cladding (see Fig. 5). The starred (*) entries in these columns correspond to IDC's and the unstarred entries to GapCap's.

TABLE II
GEOMETRICAL DETAILS OF PATTERNED METALLIC CLADDING OF METASURFACE A

| n | l | n | l | n | l | n | l |
|---|---|---|---|---|---|---|---|
| 1 | 0.237 | 41 | 0.511 | 81 | 0.519 | 121 | 0.559 |
| 2 | 0.586 | 42 | 0.484 | 82 | 0.489 | 122 | 0.534 |
| 3 | 0.623 | 43 | 0.471 | 83 | 0.456 | 123 | 0.492 |
| 4 | 0.644 | 44 | 0.465 | 84 | 0.438 | 124 | 0.451 |
| 5 | 0.633 | 45 | 0.452 | 85 | 0.451 | 125 | 0.449 |
| 6 | 0.587 | 46 | 0.417 | 86 | 0.480 | 126 | 0.497 |
| 7 | 0.468 | 47 | 0.366 | 87 | 0.500 | 127 | 0.550 |
| 8 | 0.322* | 48 | 0.318 | 88 | 0.517 | 128 | 0.588 |
| 9 | 0.535 | 49 | 0.668 | 89 | 0.669 | 129 | 0.621 |
| 10 | 1.115 | 50 | 0.955 | 90 | 0.806 | 130 | 0.663 |
| 11 | 0.947 | 51 | 0.863 | 91 | 0.792 | 131 | 0.694 |
| 12 | 0.874 | 52 | 0.780 | 92 | 0.737 | 132 | 0.688 |
| 13 | 0.849 | 53 | 0.722 | 93 | 0.682 | 133 | 0.654 |
| 14 | 0.838 | 54 | 0.681 | 94 | 0.632 | 134 | 0.607 |
| 15 | 0.828 | 55 | 0.648 | 95 | 0.593 | 135 | 0.563 |
| 16 | 0.810 | 56 | 0.620 | 96 | 0.565 | 136 | 0.538 |
| 17 | 0.780 | 57 | 0.596 | 97 | 0.552 | 137 | 0.541 |
| 18 | 0.735 | 58 | 0.576 | 98 | 0.550 | 138 | 0.562 |
| 19 | 0.675 | 59 | 0.560 | 99 | 0.550 | 139 | 0.584 |
| 20 | 0.602 | 60 | 0.543 | 100 | 0.545 | 140 | 0.593 |
| 21 | 0.525 | 61 | 0.524 | 101 | 0.526 | 141 | 0.579 |
| 22 | 0.466 | 62 | 0.497 | 102 | 0.494 | 142 | 0.539 |
| 23 | 0.444 | 63 | 0.467 | 103 | 0.456 | 143 | 0.482 |
| 24 | 0.452 | 64 | 0.441 | 104 | 0.439 | 144 | 0.447 |
| 25 | 0.489 | 65 | 0.437 | 105 | 0.468 | 145 | 0.469 |
| 26 | 0.502 | 66 | 0.448 | 106 | 0.517 | 146 | 0.524 |
| 27 | 0.395 | 67 | 0.428 | 107 | 0.550 | 147 | 0.581 |
| 28 | 0.175 | 68 | 0.401 | 108 | 0.579 | 148 | 0.626 |
| 29 | 0.647 | 69 | 0.665 | 109 | 0.666 | 149 | 0.656 |
| 30 | 1.035 | 70 | 0.876 | 110 | 0.744 | 150 | 0.666 |
| 31 | 0.891 | 71 | 0.829 | 111 | 0.744 | 151 | 0.655 |
| 32 | 0.803 | 72 | 0.759 | 112 | 0.702 | 152 | 0.620 |
| 33 | 0.756 | 73 | 0.700 | 113 | 0.650 | 153 | 0.568 |
| 34 | 0.725 | 74 | 0.653 | 114 | 0.600 | 154 | 0.517 |
| 35 | 0.701 | 75 | 0.616 | 115 | 0.563 | 155 | 0.503 |
| 36 | 0.676 | 76 | 0.588 | 116 | 0.546 | 156 | 0.540 |
| 37 | 0.648 | 77 | 0.572 | 117 | 0.548 | 157 | 0.601 |
| 38 | 0.617 | 78 | 0.561 | 118 | 0.559 | 158 | 0.650 |
| 39 | 0.582 | 79 | 0.553 | 119 | 0.566 | 159 | 0.667 |
| 40 | 0.546 | 80 | 0.540 | 120 | 0.559 | 160 | 0.620 |

ACKNOWLEDGEMENT

This work was supported by the Army Research Office under Grant W911NF-19-1-0359, and the National Science Foundation Grant Opportunities for Academic Liaison with Industry (GOALI) program under Grant 1807940.

REFERENCES

[1] J. Budhu and A. Grbic, "Recent advances in bianisotropic boundary conditions: theory, capabilities, realizations, and applications," *Nanophotonics*, vol. 10, no. 16, pp. 4075–4112, Nov. 2021.

[2] T. Brown, Y. Vahabzadeh, C. Caloz, and P. Mojabi, "Electromagnetic Inversion With Local Power Conservation for Metasurface Design," *IEEE Antennas and Wireless Propagation Letters*, vol. 19, no. 8, pp. 1291–1295, Aug. 2020.

[3] V. Ataloglou and G. Eleftheriades, "Arbitrary Wave Transformations With Huygens' Metasurfaces Through Surface-Wave Optimization," *IEEE Antennas and Wireless Propagation Letters*, vol. 20, no. 9, pp. 1750–1754, Sep. 2021.

[4] D.-H. Kwon, "Illusion electromagnetics for free-standing objects using passive lossless metasurfaces," *Physical Review B*, vol. 101, no. 23, p. 235135, Jun. 2020.

[5] J. Budhu and A. Grbic, "Fast and Accurate Optimization of Metasurfaces with Gradient Descent and the Woodbury Matrix Identity," *arXiv:2108.02762 [math.NA]*, Jul. 2021.

[6] G. Xu, S. V. Hum, and G. v. Eleftheriades, "Augmented Huygens' Metasurfaces Employing Baffles for Precise Control of Wave Transformations," *IEEE Transactions on Antennas and Propagation*, vol. 67, no. 11, pp. 6935–6946, Nov. 2019.

[7] L. Szymanski, G. Gok, and A. Grbic, "Inverse Design of Multi-input Multi-output 2D Metastructured Devices," *arXiv:2103.12210 [physics.app-ph]*, vol. 70, no. 5, pp. 3495–3505, 2021.

[8] A. Epstein and G. v. Eleftheriades, "Arbitrary power-conserving field transformations with passive lossless omega-type bianisotropic metasurfaces," *IEEE Transactions on Antennas and Propagation*, vol. 64, no. 9, pp. 3880–3895, 2016.

[9] A. Epstein and G. v. Eleftheriades, "Synthesis of Passive Lossless Metasurfaces Using Auxiliary Fields for Reflectionless Beam Splitting and Perfect Reflection," *Physical Review Letters*, vol. 117, no. 25, p. 256103, 2016.

[10] D. H. Kwon, "Lossless scalar metasurfaces for anomalous reflection based on efficient surface field optimization," *IEEE Antennas and Wireless Propagation Letters*, vol. 17, no. 7, pp. 1149–1152, 2018.

[11] D. H. Kwon, "Planar Metasurface Design for Wide-Angle Refraction Using Interface Field Optimization," *IEEE Antennas and Wireless Propagation Letters*, vol. 20, no. 4, pp. 428–432, 2021.

[12] J. Budhu, E. Michielssen, and A. Grbic, "The Design of Dual Band

Stacked Metasurfaces Using Integral Equations," *IEEE Transactions on Antennas and Propagation*, pp. 1–1, Feb. 2022.

[13] J. Budhu and A. Grbic, "Perfectly Reflecting Metasurface Reflectarrays: Mutual Coupling Modeling between Unique Elements through Homogenization," *IEEE Transactions on Antennas and Propagation*, vol. 69, no. 1, 2021.

[14] S. Pearson and S. V. Hum, "Optimization of Electromagnetic Metasurface Parameters Satisfying Far-Field Criteria," *IEEE Transactions on Antennas and Propagation*, pp. 1–1, 2021.

[15] G. Minatti, M. Faenzi, E. Martini, F. Caminita, P. de Vita, D. Gonzalez-Ovejero, M. Sabbadini, and S. Maci, "Modulated Metasurface Antennas for Space: Synthesis, Analysis and Realizations," *IEEE Transactions on Antennas and Propagation*, vol. 63, no. 4, pp. 1288–1300, Apr. 2015.

[16] A. Díaz-Rubio, V. S. Asadchy, A. Elsakka, and S. A. Tretyakov, "From the generalized reflection law to the realization of perfect anomalous reflectors," *Science Advances*, vol. 3, no. 8, Aug. 2017.

[17] H. Lee and D.-H. Kwon, "Large and efficient unidirectional plane-wave–surface-wave metasurface couplers based on modulated reactance surfaces," *Physical Review B*, vol. 103, no. 16, p. 165142, Apr. 2021.

[18] D.-H. Kwon and S. A. Tretyakov, "Perfect reflection control for impenetrable surfaces using surface waves of orthogonal polarization," *Physical Review B*, vol. 96, no. 8, p. 085438, Aug. 2017.

[19] M. Bodehou, K. al Khalifeh, S. N. Jha, and C. Craeye, "Direct Numerical Inversion Methods for the Design of Surface-Wave based Metasurface Antennas: Fundamentals, Realization, and Perspectives," *IEEE Antennas and Propagation Magazine*, vol. 64, no. 4, Aug. 2022, doi: 10.1109/MAP.2022.3169344.

[20] V. Popov, A. Díaz-Rubio, V. Asadchy, S. Tcvetkova, F. Boust, S. Tretyakov, and S. N. Burokur, "Omega-bianisotropic metasurface for converting a propagating wave into a surface wave," *Physical Review B*, vol. 100, no. 12, p. 125103, Sep. 2019.

[21] D. H. Kwon, "Modulated Reactance Surfaces for Leaky-Wave Radiation Based on Complete Aperture Field Synthesis," *IEEE Transactions on Antennas and Propagation*, vol. 68, no. 7, 2020.

[22] K. Achouri and C. Caloz, "Space-Wave Routing via Surface Waves Using a Metasurface System," *Scientific Reports*, vol. 8, no. 1, 2018.

[23] M. Almunif, J. Budhu, and A. Grbic, "Transparent, Cascaded-Sheet Metasurfaces for Field Transformations," in *IEEE International Symposium on Antennas and Propagation and USNC-URSI Radio Science Meeting (APS/URSI)*, Denver, CO, 2022, pp. 1-2.

[24] S. W. Marcus and A. Epstein, "Fabry-Pérot Huygens' metasurfaces: On homogenization of electrically thick composites," *Physical Review B*, vol. 100, no. 11. 2019.

[25] T. Brown, C. Narendra, Y. Vahabzadeh, C. Caloz, and P. Mojabi, "On the Use of Electromagnetic Inversion for Metasurface Design," *IEEE Transactions on Antennas and Propagation*, vol. 68, no. 3, pp. 1812–1824, Mar. 2020.

[26] A. Grbic and R. Merlin, "Near-Field Focusing Plates and Their Design," *IEEE Transactions on Antennas and Propagation*, vol. 56, no. 10, pp. 3159–3165, Oct. 2008.

[27] V. S. Asadchy, M. Albooyeh, S. N. Tcvetkova, A. Díaz-Rubio, Y. Ra'di, and S. A. Tretyakov, "Perfect control of reflection and refraction using spatially dispersive metasurfaces," *Physical Review B*, vol. 94, no. 7, p. 075142, Aug. 2016.

[28] D. Pozar, *Microwave Engineering*, 4th ed. Hoboken, NJ: John Wiley & Sons, 2011.

[29] A. Grbic, L. Jiang, and R. Merlin, "Near-Field Plates: Subdiffraction Focusing with Patterned Surfaces," *Science (1979)*, vol. 320, no. 5875, pp. 511–513, Apr. 2008.

[30] A. Epstein, J. P. S. Wong, and G. v. Eleftheriades, "Cavity-excited Huygens' metasurface antennas for near-unity aperture illumination efficiency from arbitrarily large apertures," *Nature Communications*, vol. 7, no. 1, p. 10360, Apr. 2016.

[31] J. M. Song and W. C. Chew, "Moment method solutions using parametric geometry," *Journal of Electromagnetic Waves and Applications*, vol. 9, no. 1–2, 1995.

[32] F. Yaman, "Numerical solution of an inverse conductive boundary value problem," *Radio Science*, vol. 43, no. 6, p. n/a-n/a, Dec. 2008.

[33] S. S. Seker and G. Apaydin, "Light scattering by thin curved dielectric surface and cylinder," in *2009 IEEE International Geoscience and Remote Sensing Symposium*, 2009, vol. 1, pp. I-29-I–32.

[34] D. Bojanjac, "Electromagnetic Wave Scattering on Planar and Cylindrical Anisotropic Structures," doctoral thesis, University of Zagreb, Faculty of Electrical Engineering and Computing, 2015.

[35] L. R. Hamilton, J. J. Ottusch, M. A. Stalzer, R. S. Turley, J. L. Visher, and S. M. Wandzura, "Numerical solution of 2-D scattering problems using high-order methods," *IEEE Transactions on Antennas and Propagation*, vol. 47, no. 4, pp. 683–691, Apr. 1999.

[36] J. Budhu and A. Grbic, "Patterned Unit Cell Design in Aperiodic Metasurfaces," in *16th International Congress on Artificial Materials for Novel Wave Phenomena – Metamaterials 2022*, Siena, Italy, 2022, pp. 1-3.

[37] J. Budhu and A. Grbic, "Unit Cell Polarizability and Sheet Impedance Extraction in Aperiodic Environments," in *2022 16th European Conference on Antennas and Propagation (EuCAP)*, Madrid, Spain, 2022.

[38] R. Barnes, "Matrix Differentiation (and some other stuff)," 2009. [Online]. Available: https://atmos.washington.edu/~dennis/MatrixCalculus.pdf. [Accessed: 28-Oct-2021].

[39] S. Wandzura, "Electric current basis functions for curved surfaces," *Electromagnetics*, vol. 12, no. 1, 1992.